\documentclass{iau}
\usepackage{graphicx}

\title[On dynamo action in the giant star Pollux : first results] 
{On dynamo action in the giant star Pollux : first results}

\author[A. Palacios \& A.S. Brun]   
{Ana Palacios$^{1,2}$
 \and Allan Sacha Brun$^2$}

\affiliation{$^1$LUPM, Universit\'e Montpellier II, \\ Place Eug\`ene Bataillon cc -0072,
F-34095, Montpellier cedex 5, France \\ email: {\tt ana.palacios@univ-montp2.fr} \\[\affilskip]
$^2$Laboratoire AIM Paris-Saclay, CEA/Irfu Universit\'e Paris-Diderot CNRS/INSU, 91191 Gif-sur-Yvette, France\\email:{\tt sacha.brun@cea.fr}}

\pubyear{2014}
\volume{302}  
\pagerange{119--126}
\jname{Title of your IAU Symposium}
\editors{A.C. Editor, B.D. Editor \& C.E. Editor, eds.}
\begin{document}

\maketitle

\begin{abstract}{
We present preliminary results of a 3D MHD simulation of the convective envelope of the giant star Pollux for which the rotation period and the magnetic field intensity have been measured from spectroscopic and spectropolarimetric observations. This giant is one of the first single giants with a detected magnetic field, and the one with the weakest field so far. Our aim is to understand the development and the action of the dynamo in its extended convective envelope. }
\end{abstract}

\firstsection 
\section{Context and numerical setup}
The detection of magnetic fields in the atmosphere of red giant stars using spectropolarimetry
(see Auri\`ere and Konstantinova-Antova in this volume) brings the opportunity to get an insight on the secular evolution of magnetic fields in solar-type and
intermediate-mass stars. Spectropolarimetry allows to estimate the magnetic fields intensity, the
doppler imaging of magnetic red giants being a delicate matter in part due to the usually long
rotation period that characterizes these objects.
The star Pollux is one of the first giants with a detected magnetic field, and its intensity of about 1 G, is the weakest detected so far. Here, we present preliminary results of a fully 3-D dynamo
computation carried out with the Anelastic Spherical Harmonic (ASH) version 2.0 code (Featherstone
et al. 2013, Brun et al. 2004), for the entire convective envelope (75\% of the total radius, strong stratification) of a 2.5 M$_\odot$ with R = 9 R$_\odot$ and L = 40 L$_\odot$, well representative of the star Pollux (see Auri\`ere et al. 2009). The hydrodynamical progenitor was computed with $\Omega = 2.63 ~10^{-7}$ rad.s$^{-1}$, corresponding to a $\upsilon$ sini $\approx 1.3$ km.s$^{-1}$, in good agreement with published values.  Using a magnetic Prandtl number of the simulation of P$_m$ = 3, we have next initialized the magnetic field with a multipole l = 3, m = 2 seed field, with an initial total magnetic energy ME less than 10$^{-3}$ of the total kinetic energy.

\begin{figure}
\includegraphics[width=0.95\textwidth]{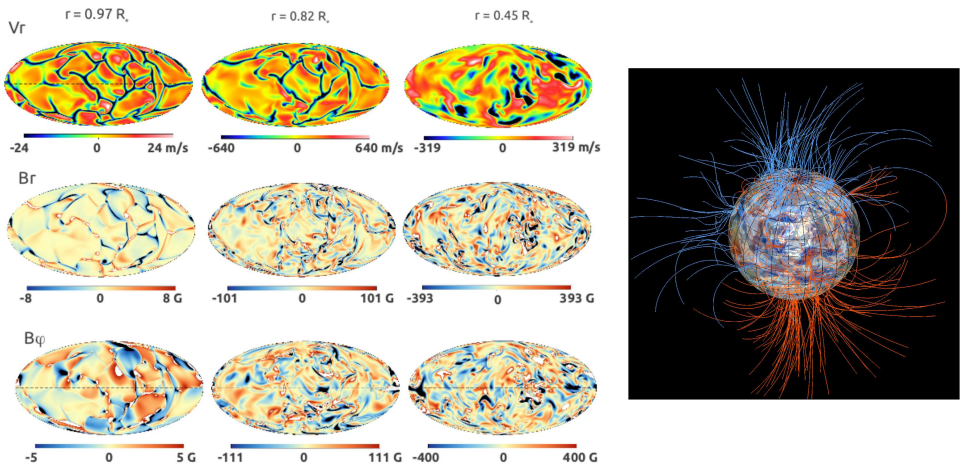}
\caption{Renderings of the simulation after evolving it for 320 days. {\underline{\sf Left} }Fluctuations of the radial velocity (convective patterns upper row), the radial magnetic field (middle row)
and the toroidal magnetic field (lower row) at three different depths. {\underline{\sf Right}} 3D rendering of the magnetic field and of the extrapolated magnetic field. The lines represent magnetic field lines and the surface is a spherical cut in the convective zone. The colours blue and orange represent negative and positive values of B$_r$.}
\end{figure}

\section{Dynamo onset, internal transport and magnetic properties.}
Our simulation is still running and as of today the dynamo is in the linear regime with an exponential growth of ME having to enter yet the saturated regime. At this point of the simulation, the total magnetic energy only corresponds to 0.1\% of the total kinetic energy. The ME is dominated by non-axisymmetric contributions of the toroidal and poloidal fields. The purely axisymmetric components TME and PME are growing but very weak. The radial energy flux balance has reached an equilibrium. The radial energy transport by the Poynting flux  is negligible (at a level of 10$^{-5}$) at this stage of the simulation. This situation should change once a saturation of the dynamo is reached. As also found in Brun \& Palacios (2009) for a more evolved red giant, the enthalpy flux dominates the energy transport. When converted to luminosity it reaches 160\% of the total stellar luminosity. This flux arises to compensate the strong inward (negative) kinetic energy flux found in the simulation.\\

Fig. 1 (left panel) shows the convective patterns achieved consisting of warm large upflows surrounded by a network of cool thin downflows. The differential rotation profile achieved in the simulation is retrograde and cylindrical in the bulk of the domain except near the rotation axis. The  rotational state is similar to that achieved in the hydrodynamical progenitor when averaged over the same period of time, indicating that the Lorentz force is too weak for the magnetic field to affect the angular velocity profile at this stage of the simulation.\\

The radial magnetic field follows the descending flows, while the toroidal field develops within that
network. The magnetic activity is larger at the bottom of the convective envelope (see scales Fig.1 left panel). The intensity of the flows is larger at larger depths, with a magnetic energy that can be locally intense but remains globally weak at this stage of the simulation. The global amplitude of the magnetic field is $\approx$ 9 Gauss. A 3D rendering of the simulation together with the extrapolated magnetic field is also shown in Fig. 1 (right panel). While the magnetic seed is a multipole and the simulation is not mature yet, we clearly see a dipolar configuration appearing (strong l=1 mode).


\end{document}